\documentclass[aps,superscriptaddress,nofootinbib,amsmath,amssymb,twocolumn,floatfix]{revtex4}

\usepackage{graphicx}
\usepackage{dcolumn}
\usepackage{hyperref}
\usepackage{bm}
\usepackage{xcolor}
\usepackage{subfigure}

\begin{document}


\title{Influence of hadronic bound states above $T_c$ on heavy-quark observables in Pb+Pb collisions at the CERN Large Hadron Collider}

\author{Marlene Nahrgang}
\email{nahrgang@subatech.in2p3.fr}
\affiliation{SUBATECH, UMR 6457, Universit\'e de Nantes, Ecole des Mines de Nantes,
IN2P3/CNRS. 4 rue Alfred Kastler, 44307 Nantes cedex 3, France }
\affiliation{Frankfurt Institute for Advanced Studies (FIAS), Ruth-Moufang-Str.~1, 60438 Frankfurt am Main, Germany}

\author{J\"org Aichelin}
\affiliation{SUBATECH, UMR 6457, Universit\'e de Nantes, Ecole des Mines de Nantes,
IN2P3/CNRS. 4 rue Alfred Kastler, 44307 Nantes cedex 3, France }

\author{Pol Bernard Gossiaux}
\affiliation{SUBATECH, UMR 6457, Universit\'e de Nantes, Ecole des Mines de Nantes,
IN2P3/CNRS. 4 rue Alfred Kastler, 44307 Nantes cedex 3, France }

\author{Klaus Werner}
\affiliation{SUBATECH, UMR 6457, Universit\'e de Nantes, Ecole des Mines de Nantes,
IN2P3/CNRS. 4 rue Alfred Kastler, 44307 Nantes cedex 3, France }

\date{\today}

\begin{abstract}
We investigate how the possible existence of hadronic bound states above the deconfinement transition temperature $T_c$ affects heavy-quark observables like the nuclear modification factor, the elliptic flow and azimuthal correlations. Lattice QCD calculations suggest that above $T_c$ the effective degrees of freedom might not be exclusively partonic but that a certain fraction of hadronic degrees of freedom might already form at higher temperatures. This is an interesting question by itself but also has a strong influence on other probes of the strongly interacting matter produced in ultrarelativistic heavy-ion collisions. A substantial fraction of hadronic bound states above $T_c$ reduces on average the interaction of the heavy quarks with colored constituents of the medium. We find that all studied observables are highly sensitive to the active degrees of freedom in the quark-gluon plasma. 
\end{abstract}


\maketitle

 \section{Introduction}

The strong suppression of high transverse momentum heavy-flavor hadrons and decay leptons in ultrarelativistic heavy-ion collisions compared to proton-proton collisions provides strong evidence for the creation of a color-deconfined plasma of quarks and gluons. It has been observed by several experiments at RHIC \cite{Phenixe,Stare,StarD} and LHC \cite{Alice,Dainese:2012ae}.
Heavy quarks are predominantly produced in the initial hard nucleon-nucleon scatterings. The subsequent interaction with the colored partonic plasma constituents leads to a substantial energy loss of high-$p_T$ heavy quarks and to partial thermalization of low-$p_T$ heavy quarks.
The in-medium energy loss is quantified by the nuclear modification factor, $R_{\rm AA}$, which is the ratio of the $p_T$-spectra measured in heavy-ion collisions and the reference $p_T$-spectra from proton-proton collisions scaled by the number of binary collisions. The $R_{\rm AA}$ of intermediate and high transverse momentum heavy-flavor hadrons and decay leptons is significantly below unity indicating that heavy quarks lose a large fraction of their initial energy traversing the partonic medium.

The finite elliptic flow, $v_2$, of heavy-flavor hadrons and decay leptons indicates that a fraction of the low- and intermediate-$p_T$ heavy quarks thermalizes in the medium and due to the scatterings in the medium, which itself exhibits an elliptic flow, starts flowing collectively with the light partons.

Most of the theoretical work on in-medium energy loss starts from perturbative QCD calculations and thus describes the interaction between heavy quarks and colored medium constituents. These models cover purely collisional energy loss scenarios \cite{Bjorken,Peshier:2006hi,Peigne:2008nd} and radiative corrections \cite{Gyulassy94,Wang95,Baier95,Baier97,Zakharov,GLV,AMY,ASW,Dokshitzer,Zhang04}. Included in numerical simulations they are mostly able to reproduce the measured $R_{\rm AA}$ by rescaling the underlying cross sections for the scattering of the heavy quarks inside the medium by a parameter $K$ \cite{Molnar:2004zj,Adare:2008cg,Gossiaux:2008jv,Uphoff:2010sh,Uphoff:2011ad,Gossiaux:2010yx,Gossiaux:2012ya} or adjusting the diffusion coefficient in a Langevin-approach \cite{Moore:2004tg,Alberico:2011zy,Cao:2012au}. This procedure is usually understood to include theoretical uncertainties attributed to effects at higher orders in perturbation theory. Naturally, one would like to compare transport coefficients resulting from the respective cross sections to lattice QCD calculations. Unfortunately, however, QCD transport coefficients cannot yet be reliably calculated on the lattice. The next best choice is to compare the results from simulations to experimental data, like the $D$ meson $R_{\rm AA}$. Here, however, one is sensitive to more than just NLO effects in pQCD calculations. In particular the details of the description of the medium affects the value of $R_{\rm AA}$ significantly at high-$p_T$ \cite{Gossiaux:2011ea}. The parameter $K$ in this context will thus in reality also depend on the special evolution of the QGP medium. 

Usually the plasma evolution is treated fluid dynamically where the equation of state at highest RHIC energies and at LHC should be dictated by lattice QCD calculations.
Here, the transition temperature $T_c$ can be obtained from a variety of observables like the chiral susceptibility, the inflection point of the energy density or the Polyakov-loop and lies in the range of $145-165$~MeV \cite{Borsanyi:2010bp}. The transition at small baryonic densities is an analytic crossover \cite{Aoki:2006we}. Around  $T_c$ lattice QCD calculations show a strong increase in thermodynamic quantities like the energy density, which is attributed to the liberation of color degrees of freedom \cite{Borsanyi:2010cj}. Below $T_c$ thermodynamic quantities agree well between lattice QCD and hadron resonance gas results. The nature of the degrees of freedom in a region directly above $T_c$ is, however, not finally answered.

In pure Yang-Mills theory, investigated in quenched lattice QCD, the Polyakov-loop is an order parameter of the $Z(3)$-symmetry and can thus be used to describe the deconfinement phase transition \cite{Pisarski:2000eq,Dumitru:2003hp}. In unquenched $2+1$ flavor lattice QCD calculations the Polyakov-loop shows a slow increase. If one defined a transition temperature from the inflection point of this increase it would be higher than those obtained from e.~g. the chiral susceptibility. In \cite{Hidaka:2008dr,Kashiwa:2013gla} the region above $T_c$ where the (renormalized) Polyakov-loop changes with temperature was called the ``semi''-QGP in contrast to higher temperatures, where the (renormalized) Polyakov-loop is flat with temperature and close to one.

In \cite{Ratti:2011au} it was recently argued that the existence of hadronic bound states above $T_c$ is very well possible, based on a comparison of  non-diagonal quark susceptibilities calculated within lattice QCD \cite{Borsanyi:2011bm} and the PNJL model \cite{Fukushima:2003fw,Ratti:2005jh} beyond mean field. In the NJL model mesons exist as resonances above $T_c$ \cite{Gastineau}.

In this work we use the Monte-Carlo propagation of heavy quarks including collisional and radiative energy loss \cite{Gossiaux:2008jv,Gossiaux:2010yx,Gossiaux:2012ya}, MC@sHQ, within a $3+1$d fluid dynamically expanding plasma given from EPOS initial conditions \cite{Werner:2010aa,Werner:2012xh}. These initial conditions are obtained from the density of relativistic strings, which describe flux tubes in multiple scattering events and fluctuate event by event. We initialize the heavy quarks randomly at the original nucleon-nucleon scattering points in these initial fluid dynamical fields according to the $p_T$-distribution from FONLL \cite{FONLL1,FONLL2,FONLL3}. Subsequent to initialization the fluid dynamical background will evolve using a parametrization of the lattice QCD equation of state \cite{Borsanyi:2010cj}. The local temperature and velocity fields then determine the scatterings of the heavy quarks with the locally thermalized partons in the plasma. The evolution of the heavy quarks is sampled by the Boltzmann equation. 
The elastic cross sections are obtained from pQCD matrix elements in Born approximation \cite{Svetitsky:1987gq,Combridge:1978kx} including a running coupling $\alpha_s$ \cite{Peshier:2006hi,Peigne:2008nd,Mattingly:1993ej,Brodsky:2002nb,Dokshitzer:1995qm} and HTL and semihard propagators, respectively \cite{Braaten,Braaten2,Gossiaux:2008jv}. 
 The incoherent emission of bremsstrahlung gluons are included via matrix elements from scalar QCD \cite{Aichelin:2013mra}. The coherent emission of gluons, i.~e. the QCD generalization of the Landau-Pomeranchuk-Migdal (LPM) effect \cite{Baier95,Baier97} is included via an effective reduction of the spectrum \cite{Gossiaux:2012cv}. 
 The hadronization of the heavy quarks will take place at the transition temperature $T_c$ via coalescence, predominantly for low-$p_T$ heavy quarks, and fragmentation, predominantly in the intermediate- and high-$p_T$ region. Previously, MC@sHQ was coupled to a $2+1$d fluid dynamical background from smooth initial conditions and an equation of state with a strong first-order phase transition \cite{Kolb:2003dz}. In the present work we will use the highly improved model to investigate the particular aspect of the nature of the degrees of freedom around the transition temperature from the perspective of heavy-quark observables.

This paper is organized as follows. In section \ref{sec:hadronicbs} we will explain how we model the fraction of hadronic bound states above $T_c$ being inspired by \cite{Ratti:2011au}. Next, we investigate the nuclear modification factor and the elliptic flow of $D$ mesons in section \ref{sec:raa} and  section \ref{sec:v2} respectively. In section \ref{sec:azimuthalcor} we look at the azimuthal correlations of $c\bar{c}$ pairs, before we conclude in section \ref{sec:con}.

 \section{Description of hadronic bound states above $T_c$}\label{sec:hadronicbs}

In \cite{Ratti:2011au} only a qualitative statement about the fraction of hadronic degrees of freedom can be made. We, therefore, need to assume some interpolation between a fully hadronic resonances gas below $T_c$ and a fully partonic medium above some certain temperature  $c\, T_c > T_c$ to quantify the mixture of degrees of freedom. 
Then the fraction of partonic degrees of freedom $\lambda$ in the medium is constructed by an exponential increase from $0$ at $T_c$ to $1$ at $c\, T_c$

\begin{equation}
 \lambda(T)=
\begin{cases}
 1 & \text{for}\quad T\geq c\, T_c \\
\exp\left(\frac{T-c\, T_c}{T-T_c}\right) & \text{for}\quad T_c<T<c\, T_c\\
 0 & \text{for}\quad T\leq T_c
\end{cases}\; .
\label{eq:lambda}
\end{equation}

Here, we choose two different scenarios with $c=1.3$ and $c=1.5$ in accordance with the conclusions drawn in \cite{Ratti:2011au}. We locate the value of $T_c$ in the middle of the range given from lattice QCD: $T_c=155$~MeV.

Since there is no strong interaction between a colored heavy quark and a color-neutral hadron we assume that a fraction of hadronic degrees of freedom above $T_c$ reduces the scattering rates for heavy quarks with the colored medium constituents. This reduction factor does not distinguish between gluons and quarks as constituents. Both scattering rates are equally multiplied by $\lambda$.

\begin{figure}
 \centering
 \includegraphics[width=0.49\textwidth]{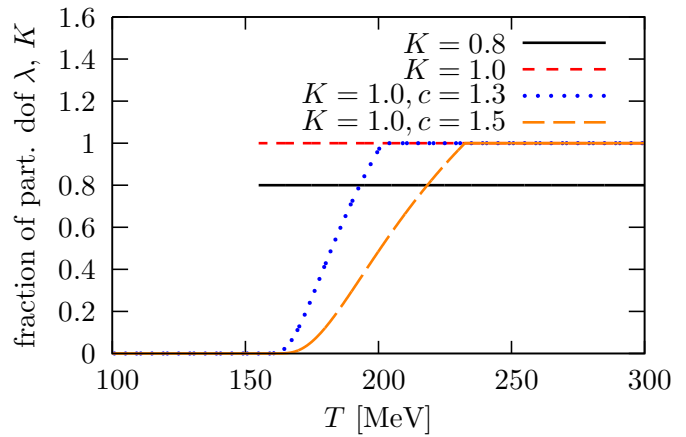}
\caption{(Color online) The fraction of partonic degrees of freedom $\lambda$ as a function of the temperature $T$. Scenarios without reduction of partonic degrees of freedom, but a global, temperature-independent rescaling by $K$ are indicated as horizontal lines, where the solid/black line is for $K=0.8$ and the dashed/red line for $K=1.0$. The two reduction scenarios are the dotted/blue line for $c=1.3$ and the orange/long-dashed line for $c=1.5$.}
\label{fig:lambda}
\end{figure}

In figure \ref{fig:lambda} the fraction of partonic degrees of freedom is shown as a function of the temperature around $T_c$. In the following sections we will compare calculations for  the two scenarios  with a reduction of partonic degrees of freedom down to $T_c$ according to equation (\ref{eq:lambda}) for $c=1.3$ and $c=1.5$ to scenarios without any reduction of partonic degrees of freedom. A scenario without reducing the partonic degrees of freedom corresponds to calculations with a global parameter $K$ independent of the temperature for $T>T_c$, which we need to adjust such that the experimental $R_{\rm AA}$ data at intermediate and high $p_T$ is best reproduced within our model. Including the fixed coupling $\alpha_s=0.3$ in the gluon emission vertex, we will see that this requires reducing the magnitude of the heavy-quark rescattering in the medium by $K<1.0$. We find that $K=0.8$ gives the best agreement with the high-$p_T$ $R_{\rm AA}$, as shown in the next section. Although a global parameter $K=1.0$, which is temperature-independent in the hot QGP phase, will lead to too much quenching and thus to too low $R_{\rm AA}$ we include this scenario in our comparison as a basis to directly see the effect of the reduction of partonic degrees of freedom. These scenarios are indicated in figure \ref{fig:lambda} by the black/solid ($K=0.8$) and dashed/red ($K=1.0$) horizontal lines. Let us repeat that a global and temperature-independent rescaling of the rates in the QGP either via the cross sections \cite{Molnar:2004zj,Adare:2008cg,Gossiaux:2008jv,Uphoff:2010sh,Uphoff:2011ad,Gossiaux:2010yx,Gossiaux:2012ya} or via the diffusion coefficient \cite{Moore:2004tg,Alberico:2011zy,Cao:2012au} is a common procedure in modeling heavy quark propagation in the QGP while our intention in this work is to investigate which influence possible hadronic bound states above $T_c$ have on heavy quark observables starting from $K=1.0$.

Several models \cite{Shuryak:2003ty,Brown:2003km,Li:2004ue} approached the possibility of the formation of bound states, colored or color-neutral, in the light quark sector in order to explain an enhanced interaction between the light partons in the medium and thus to explain the nature of the strongly-coupled fluid. For fully thermalized charm quarks the same arguments should lead to an enhanced interaction with the light partons due to the in-medium formation of $D$-meson-like resonances. For the non-equilibrium part of the heavy-flavor particle spectrum short in-medium formation times of $D$ and $B$ mesons lead to an additional dynamical quenching mechanism \cite{Adil:2006ra}. By including resonant scatterings in perturbative light-heavy parton cross section it was found that the thermalization of heavy quarks was in fact accelerated and the energy loss enhanced \cite{vanHees:2004gq,vanHees:2005wb}. We do not consider the formation of charmed hadrons inside the medium as it affects the partonic scattering cross sections only, for which in turn our assumption holds as well: If the number of partonic degrees of freedom is reduced so are the scattering rates between heavy quarks and light partons irrespectively of the precise form of the heavy-light parton scattering cross sections. 

\section{Nuclear modification factor}\label{sec:raa}

High-$p_T$ heavy quarks suffer from a strong energy loss in the initial stage of a heavy-ion collisions, where the locally thermalized medium has high temperatures and energy densities. Since the fireball created in a heavy-ion collision has a finite size these high-$p_T$ heavy quarks are expected to have left the medium and have hadronized before the bulk matter undergoes a crossover to hadronic matter. Consequently, the $R_{\rm AA}$ of high-$p_T$ heavy-flavor hadrons and decay leptons should be developed mainly during the first few fm/c of the plasma evolution and not be influenced by final hadronic interactions.

\begin{figure}
 \centering
 \includegraphics[width=0.49\textwidth]{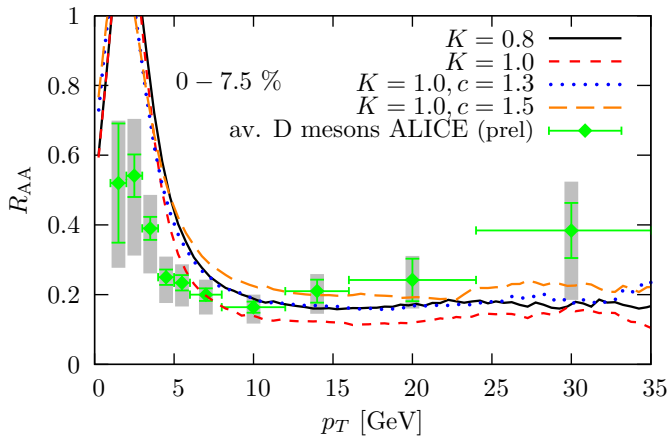}
\caption{(Color online) $R_{\rm AA}$ of D mesons at LHC in very central ($0-7.5$~\%) Pb+Pb collisions. Preliminary ALICE data of averaged $D$ mesons, including $D^0$, $D^+$ and $D^{*+}$, from \cite{delValle:2012qw}.}
\label{fig:raa}
\end{figure}

In figure \ref{fig:raa} we plot the calculated $R_{\rm AA}$ of $D$ mesons in central ($0-7.5$~\%) Pb+Pb collisions at $\sqrt{s}=2.76$~TeV. The different scenarios are compared to preliminary data from the ALICE experiment \cite{delValle:2012qw}.

First of all, we observe that a calculation without a reduction of partonic degrees of freedom above $T_c$ and $K=1.0$ shows too much suppression at $p_T>10$~GeV. This behavior in the high-$p_T$ region, on which we focus here, can be cured by a global rescaling of the rates with $K=0.8$, as mentioned in the introduction.
Applying a reduction of partonic degrees of freedom with $c=1.3$ and $c=1.5$ successively brings up the $R_{\rm AA}$ at intermediate- and high-$p_T$ as well. It is interesting to find that assuming a fraction of hadronic degrees of freedom up to $1.3\, T_c$ gives the same result for the $R_{\rm AA}$ for $p_T>7$~GeV as applying a global temperature-independent parameter of $K=0.8$.
Within the experimental errors also a larger region above $T_c$, where $\lambda<1$, can describe the high-$p_T$ data for $R_{\rm AA}$ well.

None of the scenarios investigated here reproduces the characteristic curvature in the $R_{\rm AA}$ data for $p_T>15$~GeV. This indicates that the radiative corrections including the LPM effect should be suppressed at higher $p_T$ by some further mechanism, e.~g. finite path length effects \cite{Baier97,Zakharov:2002ik,Djordjevic:2009cr} or finite life-time effects as proposed in \cite{Bluhm:2011sw}. This will, however, not change the general features of the influence of hadronic bound states above $T_c$ on heavy-quark observables, which are discussed in the following.

It is important to note that in our standard reference calculations (i.~e. without a reduction of partonic degrees of freedom) the global parameter of $K=0.8$ independent of temperature in the QGP is fixed once and for all by the description of the $R_{\rm AA}$ data. There is consequently no room to change it in the subsequent study of the elliptic flow and the azimuthal correlations (and generally also for different centralities and $B$ mesons or heavy-flavor decay electrons, which is, however, not studied in this work).
In particular the results for $K=1.0$ cannot be a valid scenario for the case without a reduction of partonic degrees of freedom and serves only as a baseline for the scenarios including a reduction of partonic degrees of freedom.

\begin{figure}
 \centering
 \includegraphics[width=0.49\textwidth]{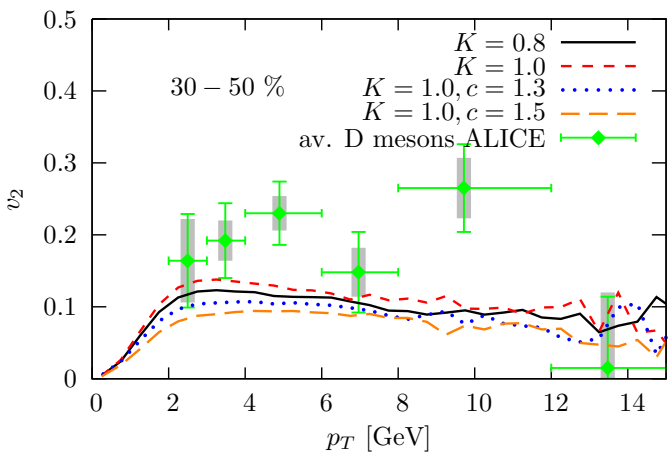}
\caption{(Color online) $v_2$ of D mesons at LHC in mid-peripheral ($30-50$~\%) Pb+Pb collisions. ALICE data of averaged $D$ mesons, including $D^0$, $D^+$ and $D^{*+}$,  from \cite{Abelev:2013lca}.}
\label{fig:v2}
\end{figure}

 \section{Elliptic flow}\label{sec:v2}

The build-up of the elliptic flow of low- and intermediate-$p_T$ heavy-flavor hadrons and decay leptons takes place predominantly at later stages of the evolution of the fireball. It is a sign for partial thermalization of the heavy quarks within the medium. The elliptic flow of the bulk matter needs to be developed first by transforming the initial elliptic shape in coordinate space into momentum space. Due to the interaction with the fluid dynamical medium the heavy quarks, which have low or intermediate $p_T$,  pick up some elliptic flow. 

Here, the situation is different than for the $R_{\rm AA}$ at higher $p_T$. While the interaction between colored heavy quarks and color-neutral hadrons is of course still negligibly small, the final hadronic interactions of low- and intermediate-$p_T$ heavy-flavor hadrons inside a hadronic medium should contribute significantly to the elliptic flow. Thus, our picture of the heavy-flavor elliptic flow remains incomplete by not including final-state hadronic interactions.

This is reflected in figure \ref{fig:v2}, where the calculated $v_2$ of $D$ mesons is shown in the $30-50$~\% centrality class from Pb+Pb collisions at $\sqrt{s}=2.76$~TeV and compared to ALICE data \cite{Abelev:2013lca}. Since our model does not cover final state interactions between heavy-flavor and light hadrons only the partonic contribution to the elliptic flow is shown. We first note that a calculation, which does not consider a reduction of partonic degrees of freedom above $T_c$ and which reproduces also the $R_{\rm AA}$ with a global rescaling of the scattering rates of $K=0.8$ also lies within the experimental uncertainties for the $v_2$ measurements.

When the scattering is enhanced by applying $K=1.0$ instead of $K=0.8$ the elliptic flow is also enhanced. But one needs to remember that this scenario is already excluded by the comparison to the $D$-meson $R_{\rm AA}$ data and is only a baseline for the scenarios where a reduction of partonic degrees of freedom is included. For both scenarios with a reduction of partonic degrees of freedom the elliptic flow is also reduced. Unlike for the $R_{\rm AA}$, however, the curves for the reduction scenario with $c=1.3$ and for the scenario without reduction of partonic degrees of freedom and $K=0.8$ do not lie on top of each other. This is a direct consequence of the fact, that the elliptic flow develops at the late stages of the evolution where the bulk matter is at temperatures above $T_c$ for which $\lambda(T)<0.8$ and thus the interaction of the charm with the collectively flowing medium is too small to transfer the flow.

A reduction of partonic degrees of freedom would hence leave room for a significant hadronic contribution, which is expected to play a role for the $D$-meson elliptic flow.

 \begin{figure}[t!]
   \subfigure[]{\label{fig:correl1}\includegraphics[width=0.44\textwidth]{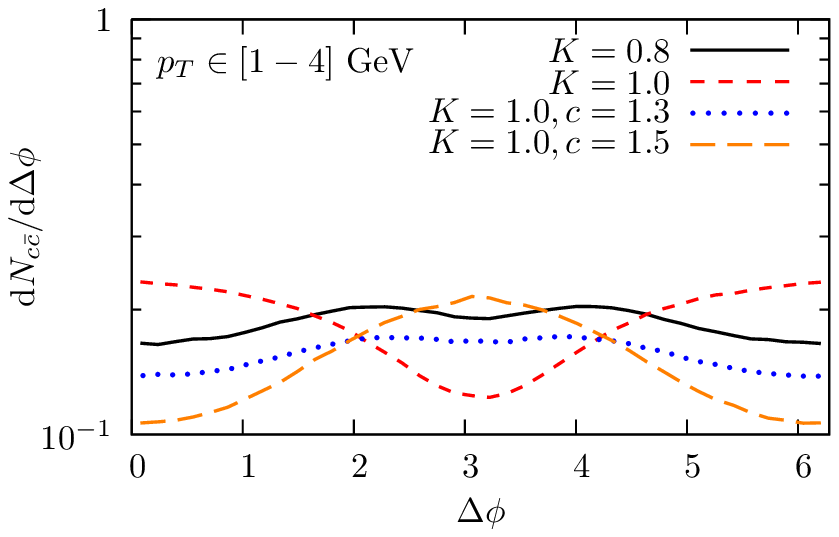}}
 
   \subfigure[]{\label{fig:correl2}\includegraphics[width=0.45\textwidth]{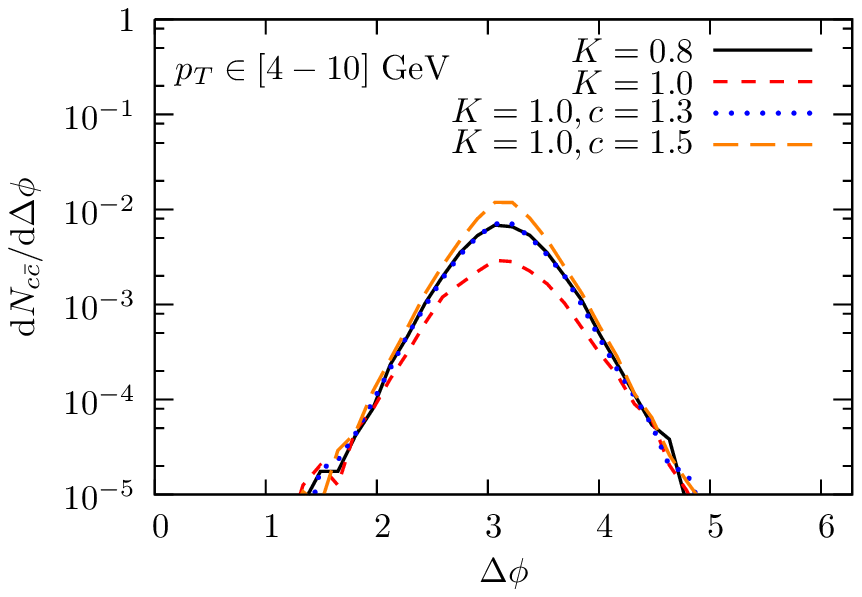}}
 
   \subfigure[]{\label{fig:correl3}\includegraphics[width=0.45\textwidth]{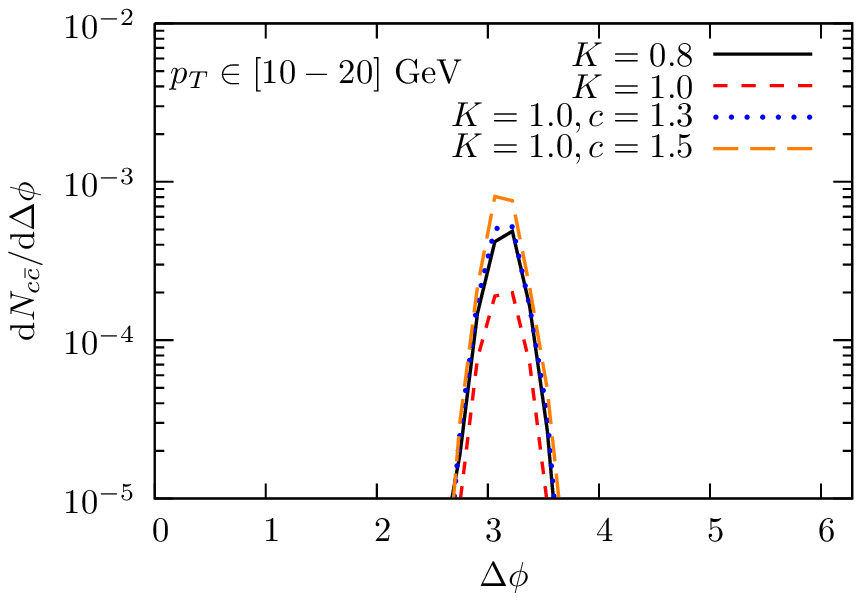}}

 \caption{(Color online) Azimuthal distribution of initially correlated $c\bar{c}$ pairs in very central ($0-7.5$\%) Pb+Pb collisions at $\sqrt{s}=2.76$~TeV. The different classes for the final charm and anticharm quark $p_T$ are  $[1-4]$~GeV \subref{fig:correl1}, $[4-10]$~GeV \subref{fig:correl2} and $[10-20]$~GeV \subref{fig:correl3}.}
 \label{fig:correl}
 \end{figure}

 \section{Azimuthal correlations}\label{sec:azimuthalcor}

Recently, we studied systematically the potential of azimuthal correlations of heavy quark-antiquark pairs to learn about the different energy loss mechanisms and their properties \cite{Nahrgang:2013saa}. Here, we investigate the azimuthal correlations of $c\bar{c}$ pairs with respect to a possible reduction of partonic degrees of freedom above $T_c$. Measurements of $D$-meson tagged azimuthal correlations should experimentally become feasible with modern detector and accelerator technologies like at the LHC. The theoretical  investigation of azimuthal correlations in heavy-ion collisions is not yet as elaborated as the study of traditional heavy-quark observables like the nuclear modification factor and the elliptic flow \cite{Zhu:2006er,Gossiaux:2006yu,Zhu:2007ne,Younus:2013be}.

As it was discussed in \cite{Nahrgang:2013saa} the final shape of the distribution of the difference of the azimuthal angle, $\Delta\phi$, of a $c\bar{c}$ pair, which was initially produced together, depends on the initialization.  Single-inclusive $p_T$ spectra for heavy-quark production in proton-proton collisions are best described by FONLL calculations \cite{FONLL1,FONLL2,FONLL3}, which does, however, not give information for more exclusive observables. The only possible leading-order production process is the annihilation of a light quark and antiquark or two gluons to form a heavy-quark pair, $q\bar{q} \to Q\bar{Q}$ or $gg \to Q\bar{Q}$. Due to energy-momentum conservation these $Q\bar{Q}$ pairs are initially correlated strictly back-to-back. Next-to-leading order (NLO) processes such as the gluon emission in the initial and final state and the production processes via flavor excitation and gluon splitting lead to a broadened peak around  $\Delta\phi=\pi$ and an additional enhancement around  $\Delta\phi=0$ in the azimuthal distributions from proton-proton collisions. Including these NLO effects one can thus find similar trends in the broadening of the back-to-back peak in the initial distribution like in the final distribution after evolution in the medium.Due to the higher mass, initial azimuthal distributions for $b\bar{b}$ pairs can be more reliably calculated than for $c\bar{c}$ pairs.
 In \cite{Nahrgang:2013saa} we were able to show that the main conclusions concerning the energy loss mechanism, the centrality and the effect of the $p_T$ trigger remains qualitatively unchanged when going from FONLL to, for example, MC@NLO \cite{Frixione:2003ei,Frixione:2002ik} initializations of $b\bar{b}$ pairs.
Here, we use the leading-order back-to-back initialization in order to study the principal effect of a reduction of partonic degrees of freedom above $T_c$ in a precise setup.

In figure \ref{fig:correl} we show the azimuthal correlations of $c\bar{c}$ pairs for central ($0-7.5$\%) Pb+Pb collisions at $\sqrt{s}=2.76$~TeV and for three different $p_T$-trigger classes: in figure \ref{fig:correl1} the final transverse momenta of the charm and the anticharm quark are between $1-4$~GeV, in figure \ref{fig:correl2} between $4-10$~GeV and in figure \ref{fig:correl3} between $10-20$~GeV.

The general trend is very well visible: Due to less rescattering the initial correlations survive the evolution in the medium more clearly in the larger $p_T$-trigger class than in the lowest $p_T$-trigger class, where the heavy quarks partially thermalize and are therefore distributed more isotropically in the azimuthal angle. For the test scenario without any reduction of partonic degrees of freedom above $T_c$ and a parameter $K=1.0$, which is however excluded by the $R_{\rm AA}$ data, the strong rescattering in the collectively flowing medium tends to azimuthally align the $c\bar{c}$ pairs instead of retaining the initial back-to-back configuration. This is the so-called ``partonic wind'' effect \cite{Zhu:2007ne}. In the more realistic scenario with a global and temperature-independent rescaling of the scattering rates in the QGP by $K=0.8$ this effect is only seen rudimentarily to the extent that a dip develops around $\Delta\phi=\pi$. For the case of a reduction of partonic degrees of freedom above $T_c$ with $c=1.3$ ($c=1.5$) this dip is much less pronounced (disappears) indicating again that effects related to flow are sensitive to the later stages of the evolution.

In all three $p_T$-trigger classes we observe that the shape of the azimuthal correlations is very similar for a global and temperature-independent rescaling in the QGP by $K=0.8$ and the scenario, where there are hadronic degrees of freedom up to $T=1.3\, T_c$. An extended region of reduced partonic degrees of freedom leads to less smearing of the initial correlations in all $p_T$-trigger classes.

 \section{Conclusions}\label{sec:con}

Within a coupled Monte-Carlo propagation of heavy quarks and a fluid dynamically evolving medium, MC@sHQ+EPOS, we studied the influence of a possible existence of hadronic bound states above the transition temperature. We assumed an exponential decrease of the partonic degrees of freedom from a fully partonic plasma at $c\, T_c$ with $c=1.3$ and $c=1.5$, and above down to a full hadronic resonance gas at $T_c$ and below. While generally the existence of hadronic degrees of freedom above $T_c$ seems likely a quantitative description is not available.

We saw that all three observables, the nuclear modification factor, the elliptic flow and the azimuthal correlations are sensitive to a possible reduction of partonic degrees of freedom above $T_c$. The value of the nuclear modification factor increases by a factor of $2$ at intermediate and larger transverse momentum when going from a scenario without any reduction of partonic degrees of freedom down to $T_c$ to our largest estimate of reduction scenarios. Due to the decreased scattering rates also the azimuthal correlations show less broadening when one opens a region of hadronic degrees of freedom above $T_c$. With these two observables it cannot be distinguished between a global temperature-independent rescaling of the scattering rates in the QGP by $K=0.8$, which is a common procedure in the numerical investigation of heavy-quark propagation, or the existence of hadronic degrees of freedom up to $T=1.3\, T_c$ given our assumptions.

Here, the elliptic flow of $D$ mesons could become more crucial. As it is built-up at the later stages of the collision a scenario with a reduction of partonic degrees of freedom up to $T=1.3\, T_c$ yields a smaller value of $v_2$ than in a scenario with a global rescaling of $K=0.8$. Since one expects a substantial contribution to the elliptic flow of $D$ mesons from final hadronic interactions ($D$ mesons interacting with light hadrons), a partonic contribution, which is itself well below the data points is most likely to be realistic. 

Our coupled model, MC@sHQ+EPOS, can of course be compared to data from RHIC experiments and additional data for heavy-flavor decay leptons at LHC energies. In a dedicated work presented elsewhere we will investigate the fluid dynamical medium from the EPOS approach, which in its current version is optimized for LHC energies, in more detail.

From our present investigation it becomes evident that a thorough understanding of the properties of the underlying bulk matter and in particular the equation of state is crucially important for understanding the heavy-quark propagation. Here, the lattice QCD equation of state requires a solid interpretation in particular with respect to the effective degrees of freedom, which are active around $T_c$. 
A solid knowledge of the underlying fluid dynamical medium turns out to be very important for a reliable study of heavy-quark observables in heavy-ion collisions.

 \section*{Acknowledgements}
We acknowledge fruitful discussions with Claudia Ratti, Rene Bellwied, Vladimir Skokov and Marcus Bluhm. This work was supported by the Hessian LOEWE initiative Helmholtz International Center for FAIR.

\end{document}